# Cosmological Constraints on $f(T, B)$ Gravity from Observations of Early and Late Universe

**Yahia Al-Omar** [1], **Majida Nahili** [1,2], and **Nidal Chamoun**[3,4]*

[1] Department of Physics, Faculty of Sciences, Damascus University, Damascus, Syria
[2] Faculty of Pharmacy, Syrian Private University, Damascus, Syria
[3] Department of Statistics, Faculty of Sciences, Damascus University, Damascus, Syria
[4] CASP, Antioch Syrian University, Maaret Saidnaya, Damascus, Syria

* Corresponding author

E-mail: yahia.alomar@damascusuniversity.edu.sy, majeda.nahili@damascusuniversity.edu.sy, and chamoun@uni-bonn.de

## Abstract

We present a unified framework that combines early- and late-Universe observations to constrain three functional realizations of $f(T, B)$ gravity: the linear, quadratic, and general power-law models. First, constraints on deviations from the standard weak interaction freeze-out temperature are derived using the most recent measurements of the primordial helium-4 mass fraction. Second, we perform a joint analysis incorporating five priors: Type Ia supernovae, baryon acoustic oscillations, cosmic chronometers, Big Bang Nucleosynthesis, and Cosmic Microwave Background in order to place bounds on the model parameters. The joint likelihood analysis significantly tightens the constraints compared to individual datasets. Third, we test the null, strong, and dominant energy conditions to evaluate the physical viability of the best-fit solutions across the cosmic redshift range. Our results show that all three $f(T, B)$ models are consistent with current observations and exhibit stable behavior under the energy-condition criteria, supporting torsion–boundary modified gravity as a robust and viable alternative of General Relativity.

# 1   Introduction

Over the past few decades, numerous independent observations have revealed a surprising picture: the universe is undergoing accelerated expansion. This conclusion began with Type Ia supernova (SNe Ia) distance measurements [1], and has since been reinforced by cosmic microwave background (CMB) anisotropies [2], baryon acoustic oscillations (BAO) [3], and large-scale structure surveys (LSS) [4].

The ΛCDM model remains the simplest framework to fit these observations. However, the lack of an explanation for dark energy and for the extremely fine-tuning of the cosmological constant motivates alternative approaches [5]. One natural option is to modify general relativity, since modified gravity models can push the acceleration back in time without invoking exotic energy components [6].

These developments arise from broader efforts to reformulate gravity using torsion rather than curvature. The Teleparallel equivalent of general relativity (TEGR) replaces curvature with torsion as the fundamental geometric object, and provides the fundamental torsion-based action. Extending TEGR to a random function $f(T)$ naturally leads to richer phenomena [7]. Furthermore, the extension $f(T, B)$, introduced in 2015 [8], combines torsion and boundary effects in a flexible framework to address both structure formation and late-time acceleration.

Big Bang nucleosynthesis (BBN) is a powerful and sensitive test of such modifications. Comparing the theoretical predictions from the modified gravity model with the observed primordial abundances imposes strong constraints on the model parameters. Recent studies have applied BBN constraints within various modified gravity framework: $f(Q)$ [9] and its extensions [10], teleparallel gravity $f(T)$ [11], the teleparallel equivalent of the Gauss-Bonnet term [12], and models incorporating the trace of the energy–momentum tensor [13]. In order to ensure consistency between BBN predictions and late-time observations, Einstein-Aether and modified Hořava–Lifshitz gravity has been examined as a gravity model using BBN and Cosmic Chronometer (CC) with fixed model parameters [14].

Recent work in the framework of $f(T, B)$ gravity has explored the theory's ability to describe the evolution of the Universe from both observational and theoretical perspectives. One study applied cosmographic methods to test the model at high redshifts, constraining the cosmographic parameters $q_0, j_0$, and $l_0$ using the Pantheon SNe Ia compilation, CC, and a newly calibrated GRB dataset [15]. Other investigations have focused on determining the Hubble parameter and estimating the best-fitting values of the model parameters using observational data from SNe Ia, CMB, BAO, and $H(z)$ [16]. The early Universe has also been studied within this framework, particularly through BBN, where several $f(T, B)$ models were tested against the observational bounds on the freeze-out temperature. Although acceptable chi-square values were obtained using fixed parameters with SNe Ia and CC data, the results did not provide the best possible fits [17]. From a theoretical standpoint, energy condition analyses have identified regions of parameter space where $f(T, B)$ models remain physically consistent [18].

Therefore, a fundamental requirement for any gravitational theory is that it must satisfy certain constraints that ensure the connection between early universe physics (such as BBN) and later observational data —such as the CMB, BAO, SNe Ia, and CC— providing a consistent description of the universe at all its epochs, and ensuring that fundamental energy conditions are met. The aim of this manuscript is just to do such an $f(T, B)$ testing study making use of the five data constraints plus energy conditions mentioned above.

This paper is organized as follows. Section 2 presents the theoretical framework for $f(T, B)$ gravity. Section 3 reviews BBN theory and observational limits on the abundance of light elements. Section 4 describes the late-time observational data, the energy conditions to be respected, and our $\chi^2$ fitting procedure. Section 5 combines BBN and the late observational data, while respecting the energy conditions, in order to test the selected $f(T, B)$ models against observations and the ΛCDM baseline. We conclude with discussion and conclusions in Section 6.

# 2 $f(T,B)$ Gravity and Cosmology

In this section, we introduce teleparallel gravity, where gravity is described by torsion rather than curvature. We start with the (TEGR) formulation [19], which uses tetrad (Vierbein) fields and the Weitzenböck connection to describe the geometry of spacetime. We then generalize this construction to $f(T)$ gravity [20]. A further generalization, $f(T,B)$ gravity, includes the boundary term $B$ allowing for a wider class of modified gravity theories [21].

In TEGR, the tetrad fields $e^A_\sigma$ connects the curved spacetime metric to the flat Minkowski metric through $g_{\sigma\tau} = \eta_{AB} e^A{}_\sigma e^B{}_\tau$, where Greek indices are used to denote spacetime coordinates and Latin indices are used to denote tangent-space components [22]. Unlike GR, which uses the torsion-free Levi-Civita connection, TEGR uses a Weitzenböck connection $\Gamma^\omega{}_{\tau\sigma}$, which has zero curvature. The gravitational effects are described by the torsion tensor $T^\omega{}_{\sigma\tau} = \Gamma^\omega{}_{\tau\sigma} - \Gamma^\omega{}_{\sigma\tau}$, and the related contorsion tensor $K^\omega{}_{\sigma\tau}$ expressing the difference between the Weitzenböck and the Levi-Civita connections. With this, the superpotential tensor is given as $S_\beta{}^{\sigma\tau} \equiv \frac{1}{2}\left(K^{\sigma\tau}{}_\beta + \delta^\sigma_\beta T^{\eta\tau}{}_\eta - \delta^\tau_\beta T^{\eta\sigma}{}_\eta\right)$, and the torsion scalar is defined as $T = S_\beta{}^{\sigma\tau} T^\beta{}_{\sigma\tau}$.

The action of TEGR is given by $S_{\text{TEGR}} = \frac{1}{16\pi G}\int d^4x\, e\, T + S_{\text{matter}}$, where $e = \det(e^A_\sigma) = \sqrt{-g}$ and $S_{\text{matter}} = \int d^4x\, e\, \mathcal{L}_m$ represents the scalar matter contribution, with $\mathcal{L}_m$ denoting the Lagrangian density of matter. By varying $S_{\text{TEGR}}$ with respect to the tetrads fields, the field equations are obtained and are mathematically equivalent to the GR equations. Generalizing TEGR allows to create a class of gravity theories. One natural generalization is by expanding $T$ to an arbitrary function $f(T)$. A more general approach unifies elements of both $f(R)$ and $f(T)$ theories by considering the function $f(T,B)$ including boundary effects [23], where the action is given as:

$$S = \frac{1}{16\pi G}\int d^4x\, e\, f(T,B) + \int d^4x\, e\, \mathcal{L}_m. \tag{1}$$

This framework demonstrates sufficient flexibility as it describes gravity in terms of torsion and/or curvature. The torsion scalar $T$ and Ricci scalar $R$ are connected by: $R = -T + \frac{2}{e}\partial_\beta(eT^\beta) = -T + B$. By varying the $f(T,B)$ action with respect to the tetrad fields, one gets the field equations containing fourth-order derivatives:

$$\begin{aligned}2e\delta^\omega_\tau \Box f_B - 2e\nabla^\omega \nabla_\tau f_B + eBf_B\delta^\omega_\tau + 4e[\partial_\sigma f_B + \partial_\sigma f_T]S_\tau{}^{\sigma\omega}\\
+ 4e^\eta_\tau \partial_\sigma(eS_\eta{}^{\sigma\omega})f_T - 4ef_T T^\beta{}_{\sigma\tau}S_\beta{}^{\omega\sigma} - ef\delta^\omega_\tau = 16\pi e\Theta^\omega_\tau,\end{aligned} \tag{2}$$

where $\Theta^\omega_\tau$ is the energy-momentum tensor, $f_T = \partial f/\partial T$, and $f_B = \partial f/\partial B$. A careful analysis of the higher-order derivatives arising from the $B$-dependence suggests that certain choices of $f(T,B)$ may introduce potential instabilities, commonly associated with Ostrogradsky-type modes [24]. Like the case of some $f(R)$ models, if one can express the fields equations in second-order form using certain properly chosen tetrad fields, then these ghosts can be avoided. However, the presence of such ghost-like degrees of freedom is model-dependent and requires detailed perturbative or Hamiltonian analysis for each specific functional form, to test whether or not the new degrees of freedom are non-dynamical and do not propagate as physical ghosts. Such an analysis lies beyond the scope of the present work.

We now apply the $f(T,B)$ gravity framework to a spatially flat FLRW universe, described by the metric $ds^2 = dt^2 - a^2(t)\delta_{ij}dx^i dx^j$, where $a(t)$ is the scale factor and $t$ is the cosmic comoving time. For this homogeneous and isotropic geometry, the tetrad field is chosen as $e_\sigma^A = \text{diag}(1, a(t), a(t), a(t))$. From this tetrad, one obtains:

$$T = -6H^2, \quad B = -6(\dot{H} + 3H^2), \tag{3}$$

where $H = \dot{a}/a$ is the Hubble parameter, and dots denote derivatives with respect to $t$. Assume the universe is filled with a perfect fluid with energy density $\rho$ and pressure $p$, the modified Friedmann equations are given by:

$$\kappa^2 \rho = -3H^2(3f_B + 2f_T) + 3H\dot{f}_B - 3\dot{H}f_B + \frac{1}{2}f, \tag{4}$$

$$\kappa^2 p = (3H^2 + \dot{H})(3f_B + 2f_T) + 2H\dot{f}_T - \ddot{f}_B - \frac{1}{2}f, \tag{5}$$

where $\kappa^2 = 8\pi G$. These equations can be formulated in the standard Friedmann form by introducing an effective dark energy component:

$$3H^2 = \kappa^2(\rho + \rho_{DE}), \quad 3H^2 + 2\dot{H} = -\kappa^2(p + p_{DE}), \tag{6}$$

where $\rho_{DE}$ ($p_{DE}$) is the effective energy density (pressure). One thus gets:

$$\kappa^2 \rho_{DE} = 3H^2(1 + 3f_B + 2f_T) - 3H\dot{f}_B + 3\dot{H}f_B - \frac{1}{2}f, \tag{7}$$

$$\kappa^2 p_{DE} = -3H^2(1 + 3f_B + 2f_T) - \dot{H}(2 + 3f_B + 2f_T) - 2H\dot{f}_T + \ddot{f}_B + \frac{1}{2}f, \tag{8}$$

This effective fluid satisfies the continuity equation $\dot{\rho}_{DE} + 3H(\rho_{DE} + p_{DE}) = 0$, which ensures energy conservation. The effective equation of state parameter can then be written as:

$$\omega_{DE} = \frac{p_{DE}}{\rho_{DE}} = -1 - \frac{\ddot{f}_B + 3\dot{H}f_B - \dot{H}(2 + 3f_B + 2f_T) - 2H\dot{f}_T - 3H\dot{f}_B}{3H^2(1 + 3f_B + 2f_T) - 3H\dot{f}_B + 3\dot{H}f_B - \frac{1}{2}f}. \tag{9}$$

When the limit $f(T,B)$ is reduced to GR with a cosmological constant, the effective equation of state parameter approaches $\omega_{DE} \to -1$, consistent with observations of late cosmological acceleration [25].

## 3 BBN Constraints from Helium and Freeze-Out Temperature

BBN occurred during the first few minutes after the Big Bang, when the temperature of the universe was about 0.1 to 1 MeV. During this brief but critical epoch, light nuclei such as helium-4, deuterium, and lithium were synthesized. The final abundances of these elements are highly sensitive to the cosmic expansion rate and, consequently, to the essential gravitational theory of the universe's dynamics. While GR has a predetermined expansionary evolution, alternative gravitational theories can modify this behavior, introducing small but detectable deviations.

During the radiation-dominated era, the universe was filled with a relativistic plasma of photons, electrons, positrons, and neutrinos in thermal equilibrium. The total energy density of the radiation is given by:

$$\rho_r = \frac{\pi^2}{30} g_* \mathcal{T}^4, \qquad (10)$$

where $g_*$ is the effective number of relativistic degrees of freedom, with $g_* \approx 10.75$ at $\mathcal{T} = 1\, \text{MeV}$ [26]. In GR, the corresponding expansion rate is $H_{GR}^2 \approx \frac{1}{3M_P^2} \rho_r$, where $M_P = 1.22 \times 10^{19}\, \text{GeV}$ is the reduced Planck mass.

Initially, the neutron-to-proton ratio remained in thermal equilibrium, maintained by weak interactions such as:

$$n + \nu_e \to p + e^-, n \to p + e^- + \bar{\nu}_e, n + e^+ \to p + \bar{\nu}_e. \qquad (11)$$

As the Universe cooled and expanded, the Hubble parameter grows faster than the reaction rate $\lambda_{tot}$, which is approximated as follows:

$$\lambda_{tot}(\mathcal{T}) \approx c_q \mathcal{T}_f^5, \qquad (12)$$

where $c_q = 9.8 \times 10^{-10}\, GeV^{-4}$ characterizes the weak interaction strength [14] (More details are provided in the appendix of [27]). The freeze-out occurred when $H = \lambda_{tot}$ at $\mathcal{T}_f \approx 8 \times 10^{-4}\, \text{GeV}$ [28]. At this point, the neutron-to-proton ratio froze at $\frac{n_n}{n_p} = e^{-Q/\mathcal{T}_f}$, where $Q = 1.293\, \text{MeV}$ is the difference in neutron and proton masses [29].

The main product of BBN is helium-4, whose primordial mass fraction depends critically on the ratio of frozen neutrons to protons. A rough estimate is given by:

$$Y_p \approx \frac{2x(t_f)}{1 + x(t_f)} e^{-(t_n - t_f)/\tau}, \qquad (13)$$

where $t_f$ is the freeze-out time, $t_n$ is the nucleosynthesis start time (set by deuterium's binding energy), and $\tau$ is the neutron lifetime ($877.75 \pm 0.28\, s$ [30]). Observations find $Y_p = 0.2449 \pm 0.0040$ [31], providing a remarkably precise benchmark for testing cosmological models. In modified gravity frameworks, the cosmic expansion rate deviates slightly from its standard GR prediction, taking the form

$$H^2 \approx \frac{1}{3M_P^2} \rho_r \left(1 + \frac{\rho_{DE}}{\rho_r}\right), \qquad (14)$$

Since $\rho_{DE} \ll \rho_r$ during the BBN, these deviations are small but not negligible. The resulting shift in the freezing temperature is

$$\frac{\delta \mathcal{T}_f}{\mathcal{T}_f} = \frac{\rho_{DE}}{\rho_r} \frac{H_{GR}}{10 c_q \mathcal{T}_f^5}; \qquad (15)$$

even this tiny correction carries forward, influencing the neutron-to-proton ratio, and the amount of helium-4 produced. The resulting change in the $Y_p$ is given by:

$$\delta Y_p = Y_p \left[ \left(1 - \frac{Y_p}{2e^{-(t_n-t_f)/\tau}}\right) \ln\left(\frac{2e^{-(t_n-t_f)/\tau}}{Y_p} - 1\right) - \frac{t_f}{\tau} \right] \frac{\delta \mathcal{T}_f}{\mathcal{T}_f}. \tag{16}$$

Remaining within the observed uncertainty ($|\delta Y_p|<0.0040$) imposes a strict constraint on the model's potential deviations

$$\left|\frac{\delta \mathcal{T}_f}{\mathcal{T}_f}\right| < 4.7 \times 10^{-4}. \tag{17}$$

This constraint represents a powerful test of modified gravity and associated cosmological models [32]. Essentially, BBNs are natural laboratory experiments that allow us to explore the laws of gravity when the universe was only a few minutes old, as described in [33].

# 4 Observational Data and Methodology

This section describes the cosmological observations and statistical methods used to constrain the parameters of our modified gravity model. We perform a full Markov chain Monte Carlo (MCMC) analysis combining information from multiple independent cosmological probes. The analysis is performed in Python, leveraging the NumPy and SciPy numerical libraries for computation. MCMC samples are sampled using emcee, and the resulting posterior distributions and confidence lines are visualized using GetDist.

## 4.1 Cosmic Chronometers (CC)

The CC method provides a model-independent and direct method for measuring the Hubble parameter, $H(z)$, over cosmological time. Unlike other cosmological probes that are based on integral values or specific cosmological assumptions, the CC method uses differential age estimates of passively evolving galaxies [34]. These galaxies, which formed their stars rapidly at high redshifts and evolved with little subsequent star formation, are excellent "cosmological clocks". Measuring the rate of change of redshift over cosmological time allows one to infer the expansion history of the universe from the equation.

$$H(z) = -\frac{1}{1+z}\frac{dz}{dt}. \tag{18}$$

In this work, we use a large set of 46 Hubble parameter observations across the redshift range $0 < z < 2.36$ [35]. These 46 CC measurements are statistically independent and are provided in the literature as uncorrelated Gaussian data points; no covariance matrix is supplied for the CC sample when constructing the likelihood. As a measure of the consistency between the theoretical predictions of our model and the observational data, we introduce the chi-squared likelihood function

$$\chi^2_{\text{CC}} = \sum_i \frac{[H_{\text{obs}}(z_i) - H_{\text{th}}(z_i)]^2}{\sigma^2_{H,i}}, \tag{19}$$

where $H_{\text{obs}}$ and $\sigma_{H,i}$ are, respectively, the observed Hubble parameter and its error at each redshift, and $H_{\text{th}}$ is the theoretical prediction from the considered cosmological model.

## 4.2 Baryon Acoustic Oscillations (BAO)

BAO's are among the most powerful cosmological probes into the history of the universe's expansion and geometry. They are echoes of sound waves that propagated through the hot photon-baryon plasma of the early universe before recombination. As photons were separated from matter, the oscillations left a distinctive signature—the sound horizon $r_d$—which serves as a "standard ruler" in the large-scale structure.

BAO observations explore this scale through the distribution of galaxies, with precise geometric constraints across the moving angular diameter distance, the Hubble distance, and the mean-scale distance:

$$D_M(z) = \int_0^z \frac{c\, dz'}{H(z')}, \quad D_H(z) = \frac{c}{H(z)}, \quad D_V(z) = [zD_H(z)D_M^2(z)]^{1/3}. \tag{20}$$

The sound horizon is defined as:

$$r_d = \int_{z_d}^\infty \frac{c_s(z)}{H(z)}\, dz, \tag{21}$$

where $z_d$ is the redshift at the baryon drag epoch, and the sound speed of the photon–baryon plasma is given by ($\Omega_b, \Omega_\gamma$ denote the density parameters for baryon, photon respectively):

$$c_s(z) = \frac{c}{\sqrt{3(1+R_b)}}, \quad R_b = \frac{3\Omega_b h^2}{4\Omega_\gamma h^2}. \tag{22}$$

This links the physics of the early universe with observations billions of years later. In our study, we use the latest DESI 2024 BAO data [36]. The DESI sample contributes 12 measurements in total: two $D_V/r_d$ measurements at $z = 0.295$ and $z = 1.491$, and ten measurements of $\{D_M/r_d, D_H/r_d\}$ at $z = 0.510, 0.706, 0.930, 1.317$, and $2.330$. The DESI covariance matrix is block-diagonal, containing correlations only between the paired $\{D_M, D_H\}$ quantities at the same redshift. We reconstruct the full $12 \times 12$ covariance matrix accordingly when computing the likelihood.

The likelihood of BAO is given as follows:

$$\chi^2_{\text{BAO}} = (\mathbf{X}_{\text{obs}} - \mathbf{X}_{\text{th}})^\top \mathbf{C}_{\text{BAO}}^{-1} (\mathbf{X}_{\text{obs}} - \mathbf{X}_{\text{th}}), \tag{23}$$

where $\mathbf{X}_{\text{obs}}$ is the vector of observed BAO measurements, $\mathbf{X}_{\text{th}}$ is the corresponding vector of theoretical predictions, and $\mathbf{C}_{\text{BAO}}$ is the full covariance matrix that takes into account statistical and systematic uncertainty [37].

## 4.3 Type Ia Supernovae (SNe Ia)

SNe Ia is a reliable tool in observational cosmology, characterized commonly as standardizable candles; these stellar explosions possess very uniform intrinsic luminosities once calibrated for light-curve shape and color. As a result of this property, they provide a direct and precise means of measuring the relation between luminosity distance and redshift, as probes of the Universe's expansion history. The theoretical distance modulus, connecting the apparent brightness of a supernova and its intrinsic luminosity, can be given as

$$\mu_{\text{th}}(z) = 5\log_{10}\left(\frac{D_L(z)}{10 \text{ pc}}\right), \tag{24}$$

where the luminosity distance $D_L(z)$ is defined by as:

$$D_L(z) = (1+z)\int_0^z \frac{c\, dz'}{H(z')}. \tag{25}$$

In this study, we utilize the Pantheon+SH0ES compilation [38], which is among the largest and most homogeneously analyzed Type Ia supernovae datasets available to date. It comprises 1701 supernovae spanning the redshift range $z < 2.3$, combining the vast Pantheon sample with the new SH0ES calibration of the absolute magnitude scale through local distance ladders. To mitigate potential systematics from local peculiar velocities and other low-redshift effects, we restrict the sample to redshifts $z < 0.01$, yielding a final sample of 1588 supernovae. Accordingly, we form a reduced data vector and a corresponding masked covariance matrix, utilizing the full covariance matrix of the compilation. The statistical tension between theory predictions and observations is then measured using a standard chi-square likelihood function defined as

$$\chi^2_{\text{SNe}} = (\boldsymbol{\mu}_{\text{obs}} - \boldsymbol{\mu}_{\text{th}})^\top \mathbf{C}^{-1}_{\text{SNe}}(\boldsymbol{\mu}_{\text{obs}} - \boldsymbol{\mu}_{\text{th}}), \tag{26}$$

where $\boldsymbol{\mu}_{\text{obs}}$ represents the observed distance moduli, $\boldsymbol{\mu}_{\text{th}}$ is the corresponding vector of theoretical distance moduli, and $\mathbf{C}_{\text{SNe}}$ is the full covariance matrix incorporating both statistical and systematic uncertainties.

### 4.4   Cosmic Microwave Background (CMB)

CMB is the residual heat from the Epoch of Recombination, when the universe was about 380,000 years old and photons decoupled from matter and have flowed freely in space ever since. In this paper, we use the 2018 Planck Compact likelihood [2], a concise summary of the important information in the full angular power spectrum of the CMB. Rather than dealing with the entire dataset, with its complex correlations between thousands of multiple poles, the compact probability reduces this to a few key cosmological parameters that are most sensitive to large-scale geometry, baryon densities, and physical matter.

The compressed parameter vector reads:

$$\mathbf{p}_{\text{CMB}} = (\Omega_b h^2,\, \Omega_m h^2,\, \theta_*), \tag{27}$$

where $\Omega_m h^2$ is the total matter density, and $\theta_*$ is the angular scale of the sound horizon at recombination. The Planck compressed likelihood provides the corresponding $3 \times 3$ covariance matrix, which incorporates both statistical uncertainties and parameter correlations obtained from the full Planck temperature and polarization power spectra. The theoretical vector includes $\theta_*^{\text{th}}$, computed following the standard Planck compressed-likelihood prescription.

The CMB data likelihood is given by

$$\chi^2_{\text{CMB}} = (\mathbf{p}_{\text{obs}} - \mathbf{p}_{\text{th}})^\top \mathbf{C}^{-1}_{\text{CMB}}(\mathbf{p}_{\text{obs}} - \mathbf{p}_{\text{th}}), \tag{28}$$

where $\mathbf{p}_{\text{obs}}$ and $\mathbf{p}_{\text{th}}$ are the observed and theoretically predicted parameter vectors, respectively. The covariance matrix $\mathbf{C}_{\text{CMB}}$ is obtained directly from the official Planck 2018 release and contains both statistical uncertainties and parameter correlations from the full energy spectrum analysis.

## 4.5 Big Bang Nucleosynthesis (BBN - Deuterium Constraint)

In addition to the helium-based BBN constraint in section 3, we include a second constraint using the primordial deuterium abundance, which provides a highly sensitive probe of the baryon-to-photon ratio. Following [2], we adopt a Gaussian prior based on the physical baryon density,

$$\Omega_b h^2 = 0.02237 \pm 0.00015. \tag{29}$$

This estimate is based on advanced simulations of nuclear reactions and high-resolution observations of deuterium absorption lines in the spectra of distant quasars. The corresponding contribution to the likelihood is

$$\chi^2_{\text{BBN}} = \frac{(\Omega_b h^2 - 0.02237)^2}{(0.00015)^2}. \tag{30}$$

The baryon density is related to the total matter density by ($c$ for "cold" dark matter) $\Omega_{m0} = \Omega_{c0} + \Omega_{b0}$, where $\Omega_{b0}$ is obtained from measuring $\Omega_b h^2$ with $h = H_0/100$. This deuterium-based constraint combines with the helium limit, for further constraining the baryon density and increasing consistency among observations of the early universe.

## 4.6 Multi-Probe Cosmological Fits

Our analysis uses a single Bayesian formula that combines the constraining power of a set of cosmic probes to test the validity of the gravitational model f(T,B). We construct a total posterior likelihood proportional to the product of the individual likelihoods for each data set. This effectively minimizes the combined $\chi^2$ statistic:

$$\chi^2_{\text{tot}} = \chi^2_{\text{CC}} + \chi^2_{\text{BBN}} + \chi^2_{\text{BAO}} + \chi^2_{\text{SNe}} + \chi^2_{\text{CMB}}. \tag{31}$$

This assumes that all data sets are statistically independent. Although this is a too-much ideal assumption, it is reasonable given that the SNe Ia, BAO, CC, BBN, and CMB data arise, each, from different physical processes and observational methods. Combining them in this way helps eliminate degeneracies between model parameters that may be difficult for individual probes to constrain individually—for example, SNe Ia only tests late cosmic expansion, while the CMB and BBN test the physics and matter content of the early universe.

In order to estimate the overall performance of the model, we use the reduced chi-square statistic:

$$\chi^2_\nu = \frac{\chi^2_{\text{tot}}}{N_{\text{data}} - N_{\text{param}}}, \tag{32}$$

where $N_{\text{data}}$ is the number of data points and $N_{\text{param}}$ the number of fitted parameters. A value of $\chi^2_\nu \approx 1$ is a good fit, while large deviations indicate model flaws or over/underestimated uncertainty. Therefore, this Bayesian multi-probe analysis constitutes a robust and balanced test of the $f(T, B)$ model over cosmic time [39–41].

## 4.7 Energy conditions

Energy conditions lie at the core of any gravitational theory because they impose general physical constraints on the energy-momentum tensor to make matter plausible. In GR, they form the basis of important results such as singularity theories and that gravity is attractive. In modified gravity theories, energy conditions serve as a good tool for verifying the physical plausibility of effective energy-momentum components arising from geometric modifications [42].

For a perfect fluid with energy density and isotropic pressure, the energy–momentum tensor is ($u_\mu$ is the velocity 4-vector):

$$T_{\mu\nu} = (\rho + p)u_\mu u_\nu + p g_{\mu\nu}, \qquad (33)$$

In modified gravity models, such as $f(T, B)$, the geometric modifications can be interpreted as an effective fluid with energy density and pressure.

$$\rho_{eff} = \rho + \rho_{DE}, \qquad p_{eff} = p + p_{DE}. \qquad (34)$$

The standard energy conditions then apply to these effective quantities:

- NEC (Null Energy Condition): $\rho_{eff} + p_{eff} \geq 0$,
- DEC (Dominant Energy Condition): $\rho_{eff} \geq |p_{eff}|$,
- SEC (Strong Energy Condition): $\rho_{eff} + 3p_{eff} \geq 0$, $\rho_{eff} + p_{eff} \geq 0$.

All of these conditions have clear physical meanings: DEC ensures causal energy flow, with energy density dominating over pressure, preventing superluminal propagation of matter and energy. NEC prevents negative energy transfer along zero-order curves, and SEC guarantees attractive gravity (violation of which allows cosmic acceleration) [43]. Violations of these assumptions highlight the influence of the geometric sector. For example, a violation of SEC (<0) corresponds to repulsive gravity, which could lead to a cosmic acceleration at a later time. Violations of NEC or DEC indicate the presence of exotic matter or energy, which could lead to unconventional or non-standard physical phenomena.

# 5 Models

## 5.1 The Linear Model

Inspired by the simplest combination of torsion and boundary terms, as suggested by the Noether-symmetry approach [44], we begin with a linear extension of TEGR. This motivates the Linear model:

$$f(T, B) = \alpha T + \gamma B, \qquad (35)$$

where $\alpha$ and $\gamma$ are dimensionless parameters. This framework elegantly bridges between well-known gravitational theories through specific parameter choices: for $\gamma = 0$, the model reduces to TEGR, effectively rescaling the gravitational constant to $G_{\text{eff}} = G/\alpha$; when $\alpha = -\gamma$, it simplifies to $f(R) = \gamma R$, recovering GR with $G_{\text{eff}} = G/\gamma$; whereas putting $\alpha = \gamma = 1$ restores standard GR. One can obtain $\alpha$ by using the functional form (Eq. 35) in the first modified Friedmann equation (Eq. 4) evaluated at the present time:

$$\alpha = -(\Omega_{m0} + \Omega_{r0} + 5\gamma)/3, \qquad (36)$$

Substituting Eq. (35) in Eq. (7), and defining $\Omega_{F0} = 1 - \Omega_{m0} - \Omega_{r0}$, one obtains $\rho_{DE}$ as:

$$\rho_{DE} = \rho_r[\Omega_{F0} - 3\gamma], \qquad (37)$$

Likewise, substituting Eq. (37) along with Eq. (10) into Eq. (15) we get

$$\frac{\delta \mathcal{T}_f}{\mathcal{T}_f} = \frac{\sqrt{\pi^2 g_*}(\Omega_{F0} - 3\gamma)}{30\sqrt{10} q M_P \mathcal{T}_f^3}, \tag{38}$$

BBN imposes stringent bounds on the allowed deviation in the freeze-out temperature $\mathcal{T}_f$, thereby constraining the parameter $\gamma$. As illustrated in Fig. 1, consistency with the observed primordial abundances of light elements requires $0.229 \leq \gamma \leq 0.237$.

In our analysis, we adopt fixed values for cosmological parameters: the present-day density parameters for matter and radiation are set to $\Omega_{m0} = 0.315$ and $\Omega_{r0} = 8.4 \times 10^{-5}$, respectively [2], whereas the freeze-out temperature was set to $\mathcal{T}_f = 6 \times 10^{-4}$ GeV [11]. Aside from $\Omega_{m0}$ which will be treated as a free parameter in the remaining analysis, the other parameters values, in addition to the present CMB temperature fixed at $\mathcal{T}_0 = 2.6 \times 10^{-13}$ GeV [45], are consistently maintained across all models examined in this study.

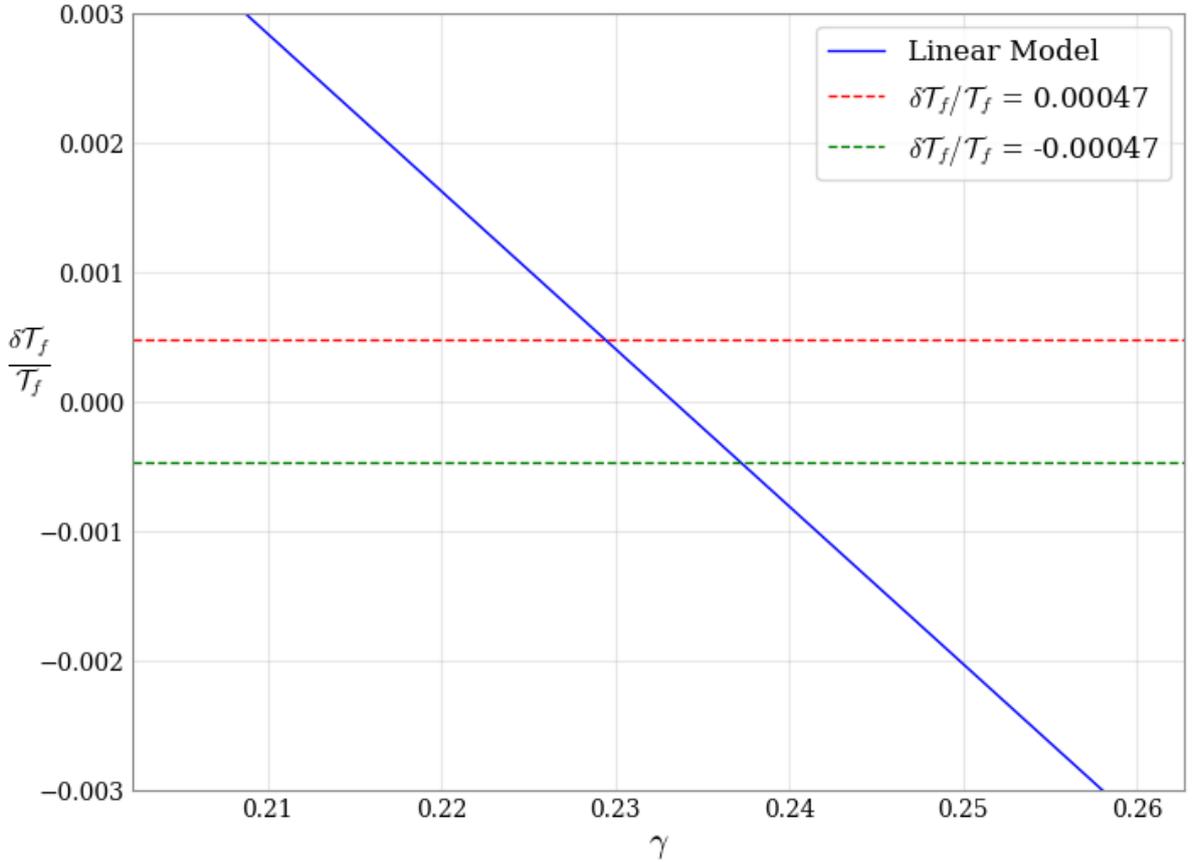

**Figure 1.** The relationship between $\delta \mathcal{T}_f / \mathcal{T}_f$, and $\gamma$ shows that the rules set by BBN restrict $\gamma$ to values roughly between 0.229 and 0.237.

Substituting the functional form of the linear model and the parameter $\alpha$ into the first Friedmann equation leads to the following differential equation for the dimensionless Hubble parameter $E(z) = H(z)/H_0$:

$$\frac{dE(z)}{dz} = \frac{1}{2\gamma E(z)} \left[ (\Omega_{m0}(1+z)^2 + \Omega_{r0}(1+z)^3) + \frac{(\gamma - \Omega_{m0} - \Omega_{r0})}{(1+z)} E(z) \right], \tag{39}$$

For the linear model, cosmological constraints from individual and combined datasets are presented in Table 1. Eq. 39 is used in Fig. 2 showing the corner plots of the marginalized posterior distributions from different dataset combinations, and in Fig.3 and Fig 4 corresponding to individual data sets.

We find that the matter density parameter $\Omega_{m0}$ varies in the range of $0.30 - 0.36$, with individual probes such as CC (Fig. 3) and SNe Ia (Fig. 4) favoring slightly higher values. However, a combined analysis with the CMB gives $\Omega_{m0} = 0.303 \pm 0.003$, consistent with the ΛCDM. The fitted value of the parameter ($\gamma$) in individual probes is given with a high uncertainty in the range of (0.46–0.52), but converges to $0.386 \pm 0.016$ when all datasets are combined. Similarly, $H_0$ ranges from ~71 km s$^{-1}$Mpc$^{-1}$ for SNe Ia (Fig. 4) to a constant ~68.7 km s$^{-1}$Mpc$^{-1}$ in the full fit.

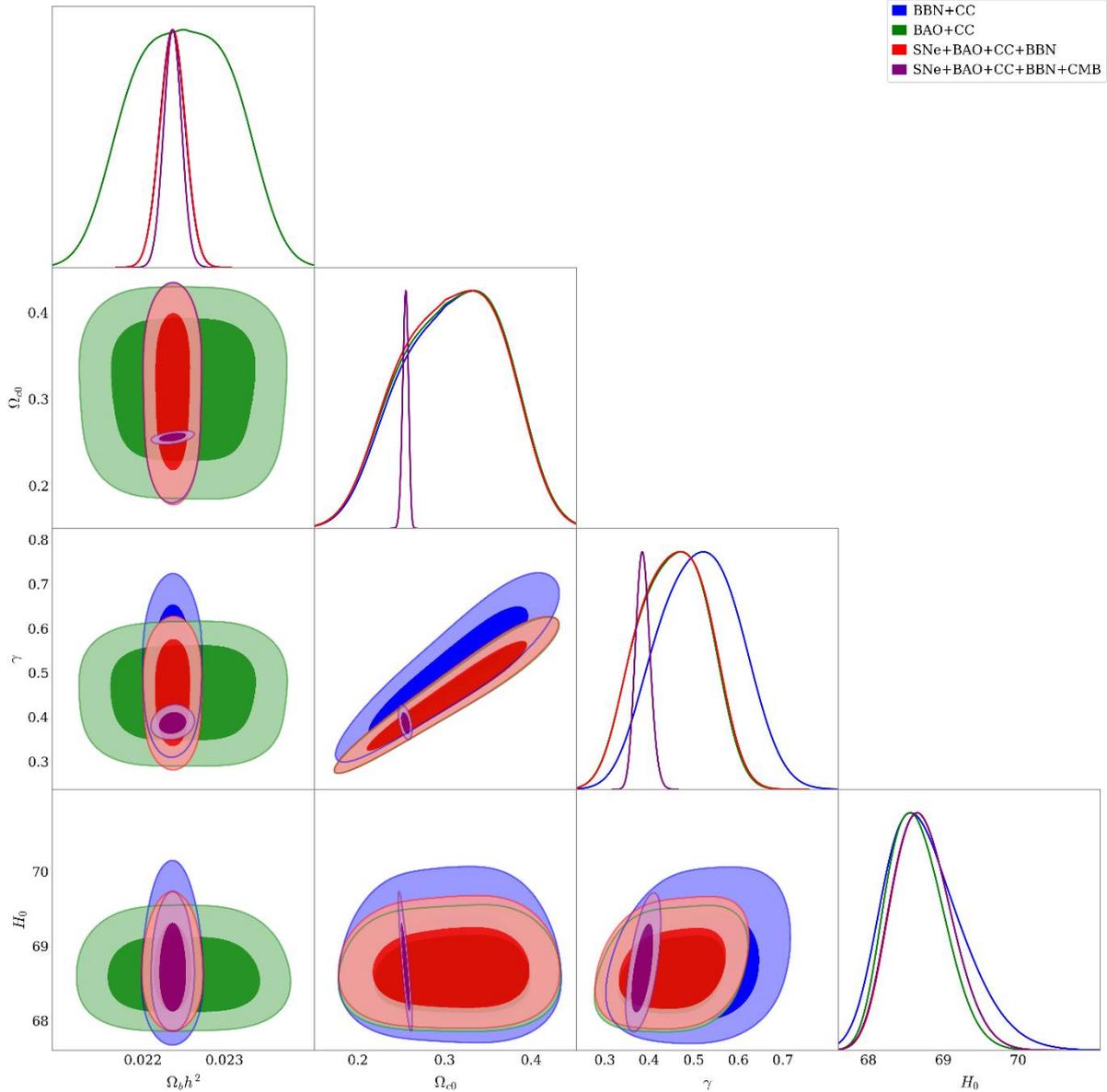

**Figure 2.** Corner plot showing the marginalized posterior distributions of the linear model parameters from different dataset combinations. The 68% and 95% confidence regions highlight the consistency among probes and the tightening of constraints in the analysis.

Despite this close agreement with the late-time data, the best fit (γ) remains well above the BBN helium limit ($0.229 \leq \gamma \leq 0.237$), indicating a persistent tension between early nucleosynthesis and late-time cosmology. However, in the joint fit (SNe Ia + BAO + CC + BBN + CMB), all energy conditions (NEC, DEC, SEC) are satisfied, demonstrating that the effective cosmic fluid in the $f(T, B)$ scenario remains physically sensitive and phantom-free.

**Table 1.** Best-fitting parameters for the linear model from different cosmological data sets, with 68% confidence intervals and $\chi_\nu^2$ values.

| Dataset | $\Omega_{m0}$ | $\gamma$ | $H_0$ | $\chi_\nu^2$ |
| --- | --- | --- | --- | --- |
| SNe | $0.3494^{+0.0996}_{-0.0838}$ | $0.5227^{+0.3170}_{-0.2964}$ | $71.15^{+1.28}_{-1.67}$ | 0.2 |
| CC | $0.3644^{+0.0927}_{-0.0867}$ | $0.5182^{+0.0896}_{-0.1027}$ | $68.64^{+0.59}_{-0.41}$ | 0.54 |

| Dataset | $\Omega_{m0}$ | $\gamma$ | $H_0$ | $\chi^2_\nu$ |
|---|---|---|---|---|
| BBN+CC | $0.3641^{+0.0909}_{-0.0858}$ | $0.5167^{+0.0887}_{-0.1017}$ | $68.67^{+0.59}_{-0.45}$ | 0.52 |
| BAO+CC | $0.3631^{+0.0919}_{-0.0861}$ | $0.4585^{+0.0767}_{-0.0901}$ | $68.60^{+0.41}_{-0.36}$ | 0.73 |
| SNe+BAO+CC+BBN | $0.3601^{+0.0911}_{-0.0841}$ | $0.4588^{+0.0784}_{-0.0898}$ | $68.69^{+0.41}_{-0.37}$ | 0.21 |
| SNe+BAO+CC+BBN+CMB | $0.3031^{+0.0031}_{-0.0034}$ | $0.3859^{+0.0162}_{-0.0152}$ | $68.69^{+0.42}_{-0.38}$ | 0.22 |

Inserting the best-fit and BBN-constrained values of $\gamma$, and $\Omega_{m0}$ into (Eq. 36), one obtains $\alpha = -0.488$ (BBN), and $\alpha = -0.733$ (joint fit).

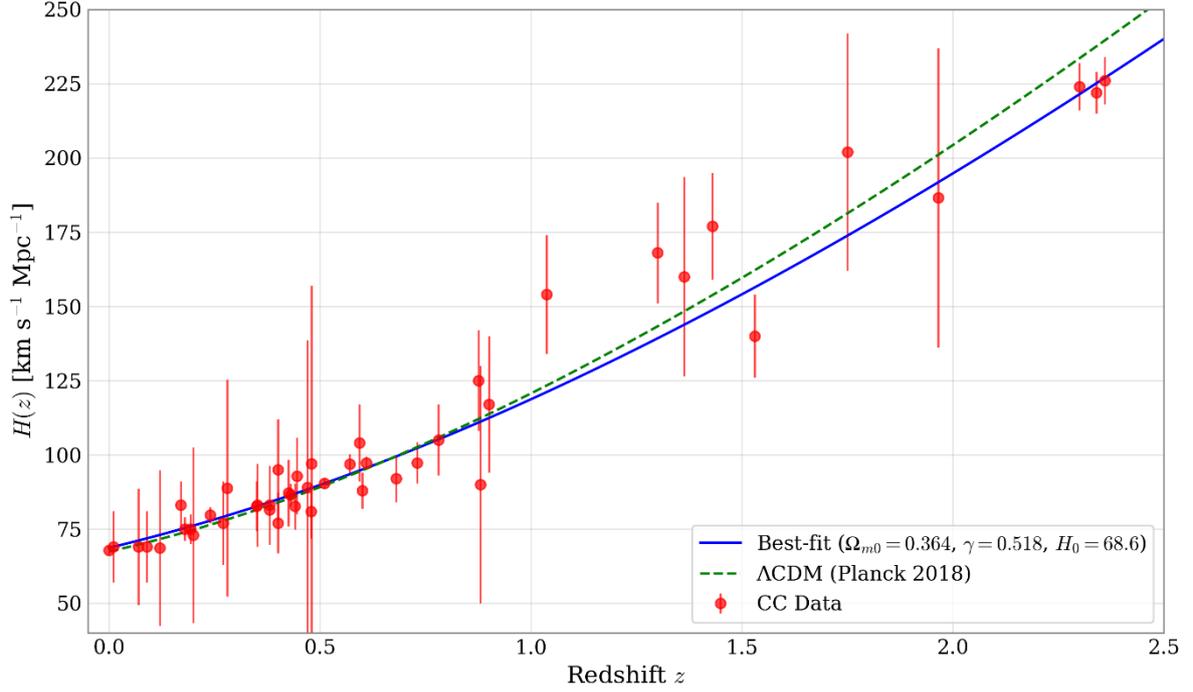

**Figure 3.** Best-fit evolution of the $H(z)$ for the linear model using CC data. The solid curve corresponds to the model prediction, while the dashed line represents the ΛCDM reference from Planck 2018.

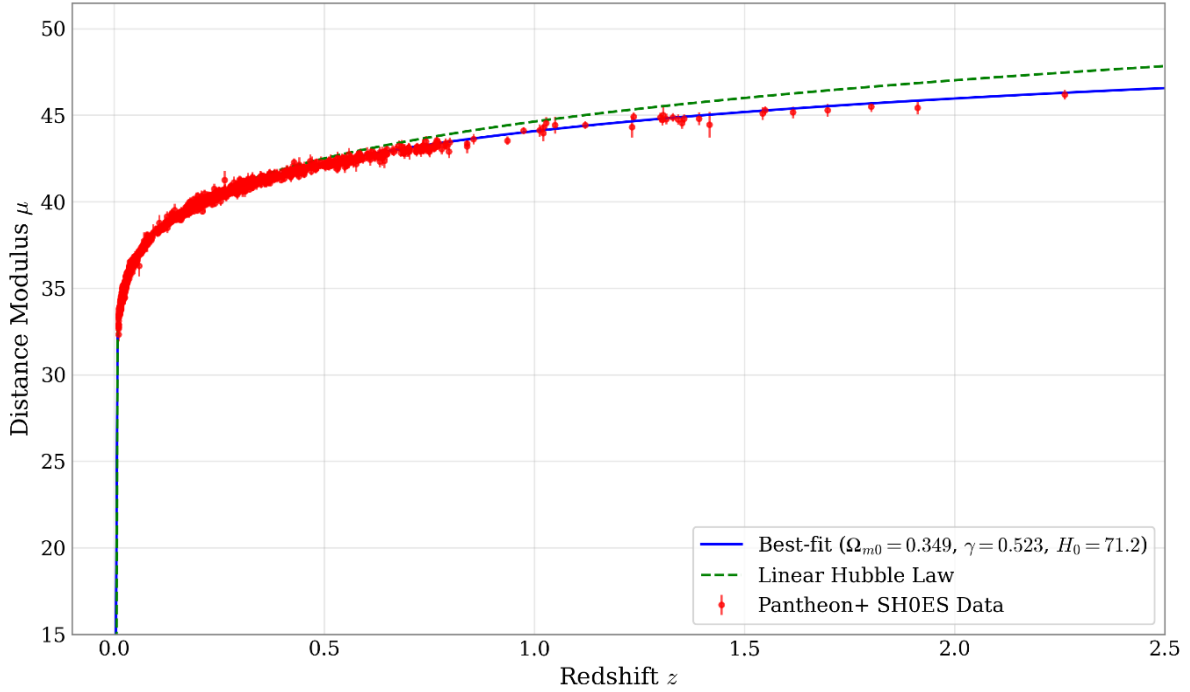

**Figure 4.** Distance modulus–redshift relation for the linear model constrained by the Pantheon+SH0ES supernova sample. The solid curve depicts the best-fit model prediction, and the dashed line indicates the linear Hubble law.

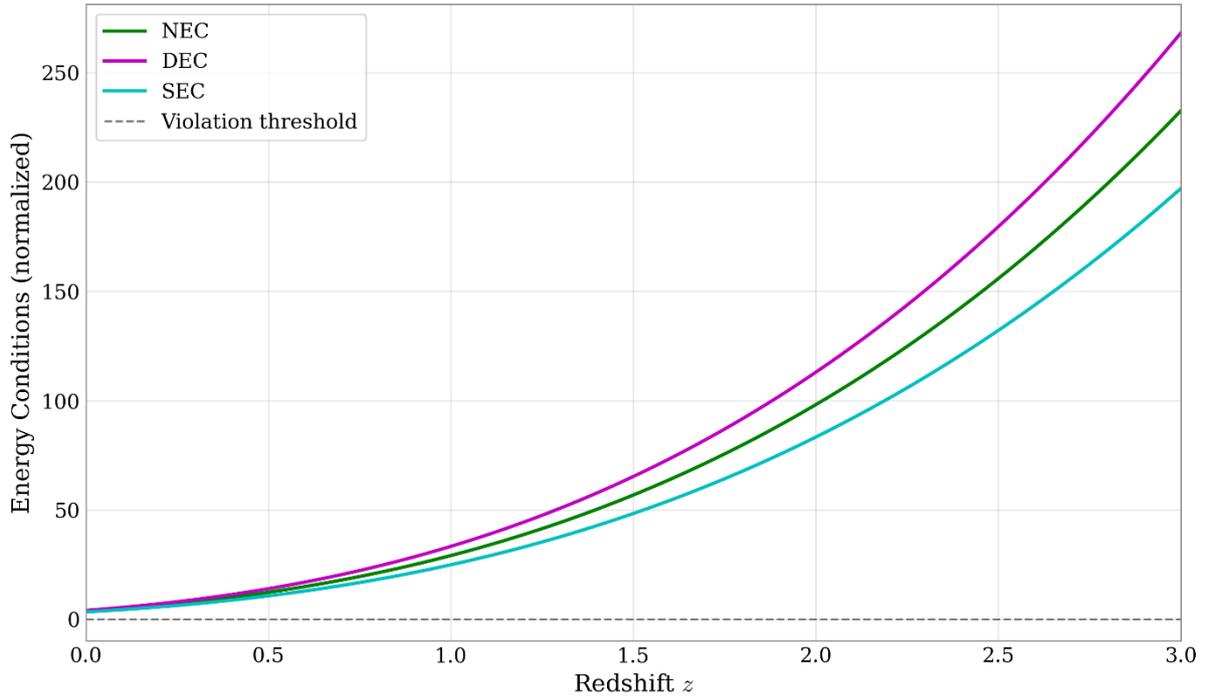

**Figure 5.** Evolution of the normalized energy conditions (NEC, DEC, and SEC) as functions of redshift $z$ for the linear model. The dashed line marks the violation threshold, indicating the limits where the energy conditions cease to hold.

## 5.2 The Square Model

To explore minimal polynomial deviations beyond the linear torsion structure of TEGR, we are inspired by previous studies in modified gravity frameworks, such as $f(T,\mathcal{T})$ gravity, where a quadratic torsion term was considered to investigate early- and late-time cosmological dynamics [46]. Following this idea while adapting it to $f(T,B)$, we introduce a quadratic torsion term while retaining the boundary term linear, leading to the Square model:

$$f(T,B) = -T + \varepsilon T^2 + \lambda B, \tag{40}$$

where $\lambda$ and $\tilde{\varepsilon}$ are dimensionless couplings, $\varepsilon = \tilde{\varepsilon}/M^2$ (where $M$ is a reference energy scale, e.g., the Planck mass). Setting $\lambda = 0$ yields a pure $f(T)$ theory, and further choosing $\varepsilon = 0$ reproduces TEGR (up to an overall sign convention absorbed into the action's normalization), with $|G_{eff}| = G$. Conversely, taking $\varepsilon = 0$ and $\lambda = 1$ gives $f(T,B) = R$, thereby recovering GR with $G_{eff} = G$.

One can obtain $\varepsilon$ by plugging this model form (Eq. 40) into (Eq. 4) at the present time:

$$\varepsilon = [\lambda + 1/5(\Omega_{m0} + \Omega_{r0}) - 3/5]/6H_0^2, \tag{41}$$

Substituting (Eq. 40) into (Eq. 7), one obtains $\rho_{DE}$ as:

$$\rho_{DE} = \rho_r[2\lambda - 2 + (\Omega_{F0} + 2 - 5\lambda)H^2/H_0^2], \tag{42}$$

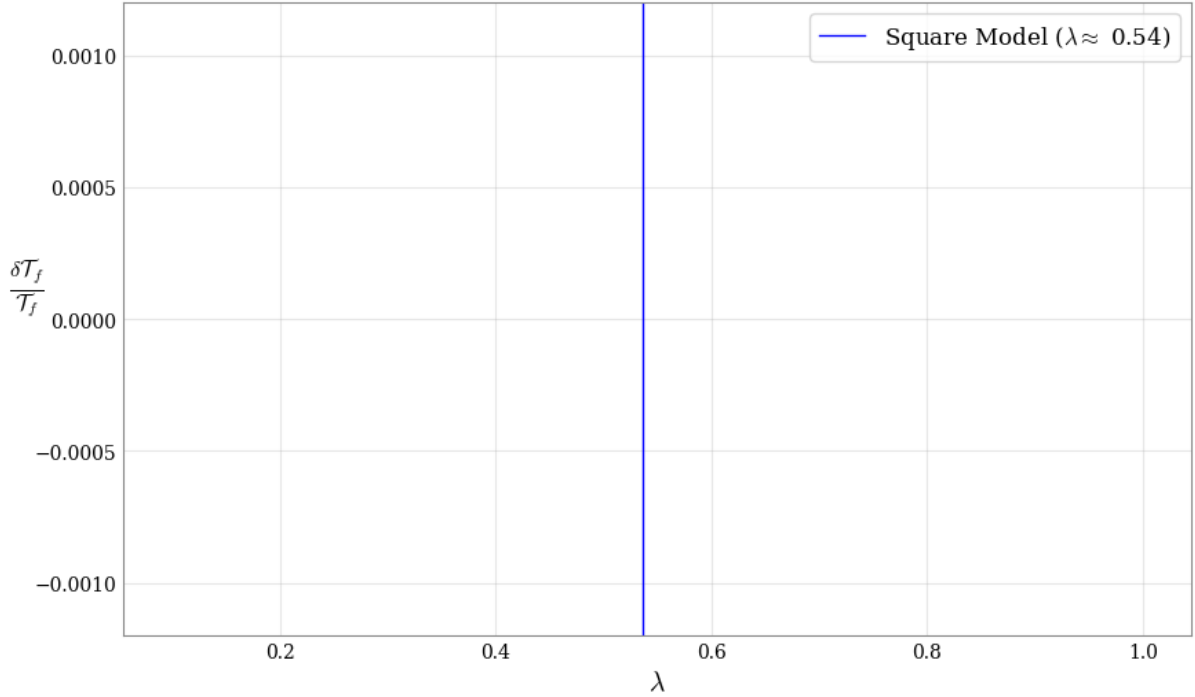

**Figure 6.** The square model passes the BBN constraints for $\lambda \approx 0.54$

Putting $\rho_{DE}$ (Eq. 42), and (Eq. 10) in (Eq. 15), we obtain:

$$\frac{\delta \mathcal{T}_f}{\mathcal{T}_f} = \sqrt{\frac{\pi^2 g_*}{90}\left[2\lambda - 2 + (\Omega_{F0} + 2 - 5\lambda)\left(\frac{\mathcal{T}_f}{\mathcal{T}_0}\right)^4\right]}\frac{1}{10qM_P\mathcal{T}_f^3}. \tag{43}$$

Compatibility with BBN constraints is achieved when $\lambda = 0.53698$, as Fig. 6 shows. Actually the allowed BBN limits would make $\lambda$ rather constant varying in a very slim interval.

**Table 2.** Best-fitting parameters for the square model from different cosmological data sets, with 68% confidence intervals and $\chi_\nu^2$ values.

| Dataset | $\Omega_{m0}$ | $\lambda$ | $H_0$ | $\chi_\nu^2$ |
|---|---|---|---|---|
| SNe | $0.3485^{+0.0992}_{-0.0821}$ | $0.5450^{+0.0355}_{-0.0325}$ | $71.27^{+1.21}_{-1.69}$ | 0.2 |
| CC | $0.3895^{+0.0407}_{-0.0621}$ | $0.5246^{+0.0069}_{-0.0052}$ | $68.63^{+0.58}_{-0.42}$ | 0.53 |
| BBN+CC | $0.3829^{+0.0443}_{-0.0612}$ | $0.5254^{+0.0068}_{-0.0055}$ | $68.68^{+0.57}_{-0.45}$ | 0.52 |
| BAO+CC | $0.4023^{+0.0314}_{-0.0444}$ | $0.5192^{+0.0056}_{-0.0040}$ | $68.73^{+0.48}_{-0.41}$ | 0.74 |
| SNe+BAO+CC+BBN | $0.4007^{+0.0319}_{-0.0422}$ | $0.5195^{+0.0052}_{-0.0042}$ | $68.79^{+0.46}_{-0.41}$ | 0.21 |
| SNe+BAO+CC+BBN+CMB | $0.2954^{+0.0031}_{-0.0030}$ | $0.5312^{+0.0015}_{-0.0014}$ | $69.59^{+0.36}_{-0.36}$ | 0.22 |

Substituting (Eq.40) along with (Eq. 41) in (Eq.4) yields:

$$\frac{dE(z)}{dz} = \frac{1}{2\lambda(1+z)} \Big[ (\Omega_{m0}(1+z)^3 + \Omega_{r0}(1+z)^4)E^{-1}(z) + (6\lambda - 3)E^2(z) - \big((\Omega_{m0} + \Omega_{r0}) + 5\lambda - 3\big)E^3(z) \Big], \quad (44)$$

For the square model, the full cosmological constraints are given in Table 2. Eq. 44 was used for the fitting shown in Figs. 7,8 and 9.

The matter density parameter $\Omega_{m0}$ ranges from 0.295 to 0.40, with individual probes such as CC (Fig. 8) and SNe Ia (Figure 9) favoring slightly higher values. When the CMB datasets are included, $\Omega_{m0}$ is fixed to 0.295 ± 0.003, highlighting the strong constraining power of the combined datasets. The gravitational coefficient $\lambda$ is also consistent through the analyses, converging to $\lambda = 0.5312 \pm 0.0015$ in the combined fit, while it can crucially implement the helium BBN constraint at $\lambda = 0.5369$, demonstrating consistency between early and late cosmology. The Hubble constant also lies within $68.6 - 71.3 \text{ km s}^{-1}\text{Mpc}^{-1}$, maintaining internal consistency.

For the combined case (SNe Ia + BAO + CC + BBN + CMB), the energy condition leads to Fig. 10, which shows that NEC and DEC are satisfied, while SEC is violated—indicating repulsive gravitational effects consistent with the observed cosmic acceleration. Inserting the best fit $\lambda$ into (Eq. 41) with $(M = H_0)$ yields a dimensionless parameter $\tilde{\varepsilon} = 0.0783$.

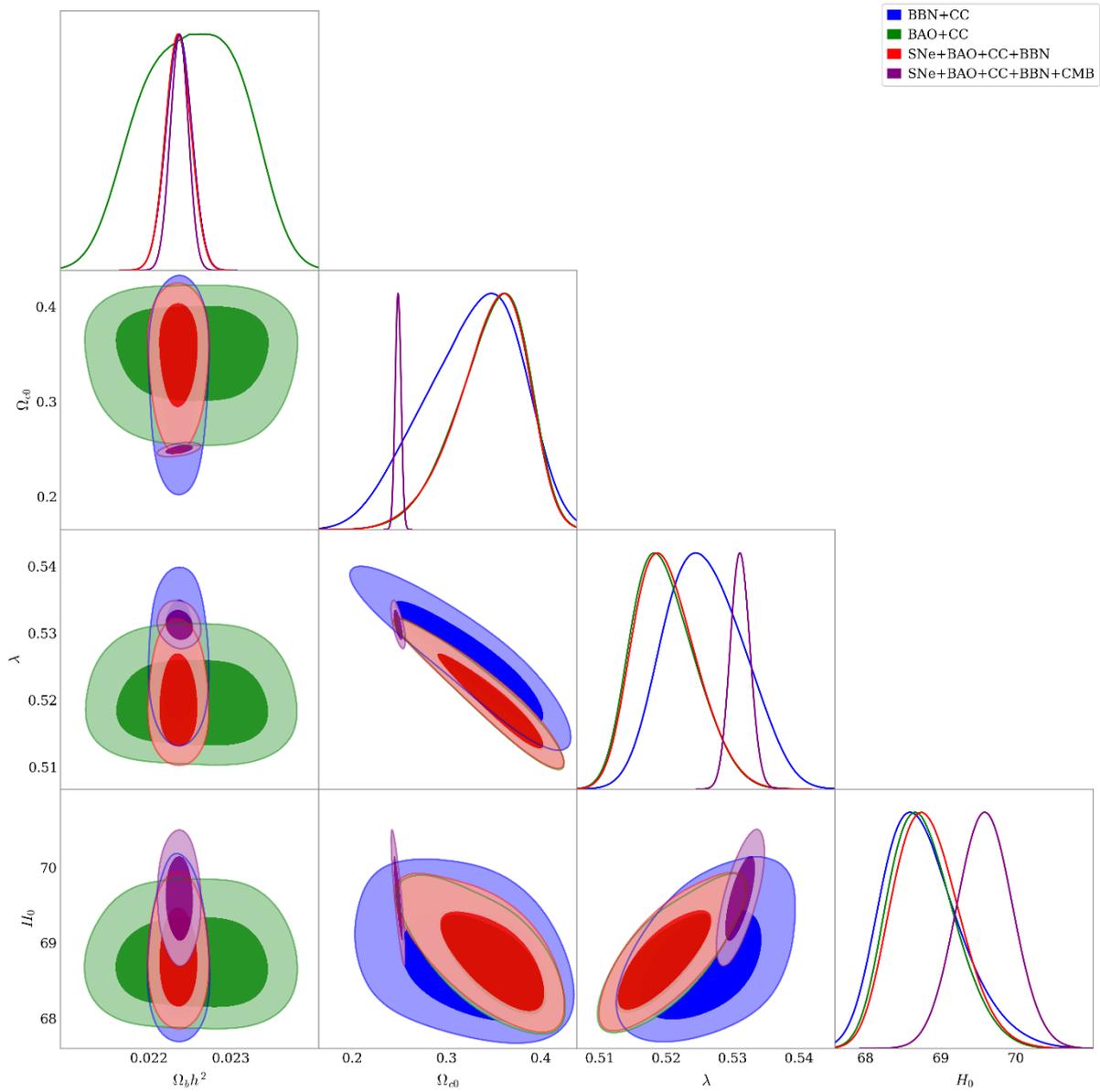

**Figure 7.** Corner plot showing the joint posterior distributions of the square model parameters derived from different observational datasets, showing 68% and 95% confidence regions.

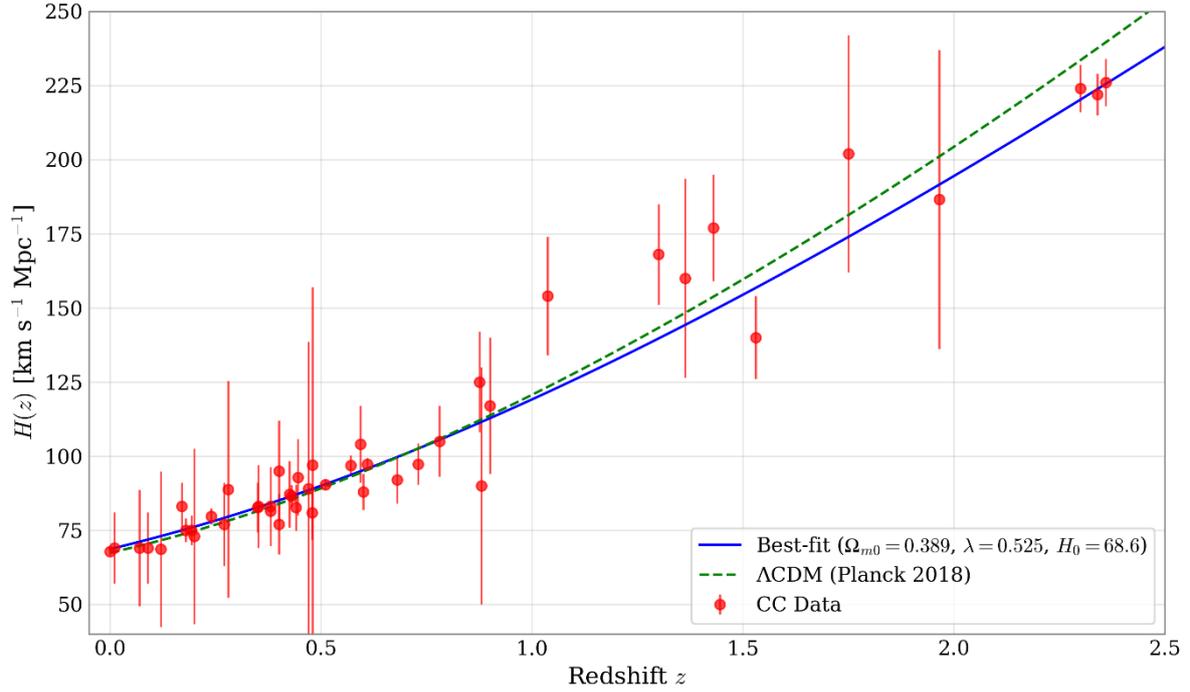

**Figure 8.** Reconstructed Hubble parameter evolution for the square model using CC data. The solid curve represents the best-fit model, while the dashed line corresponds to the ΛCDM prediction from Planck 2018.

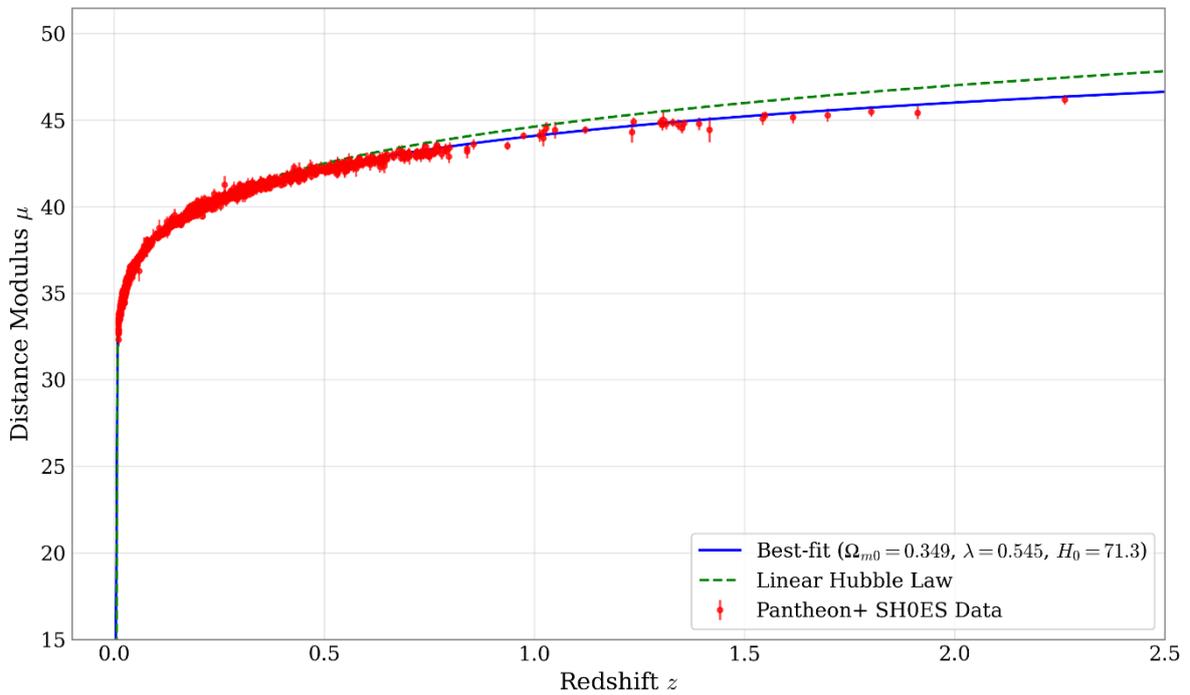

**Figure 9.** Distance modulus–redshift relation for the square model fitted to Pantheon+SH0ES Type Ia supernova data. The solid line denotes the best-fit theoretical prediction, while the dashed line represents the linear Hubble law.

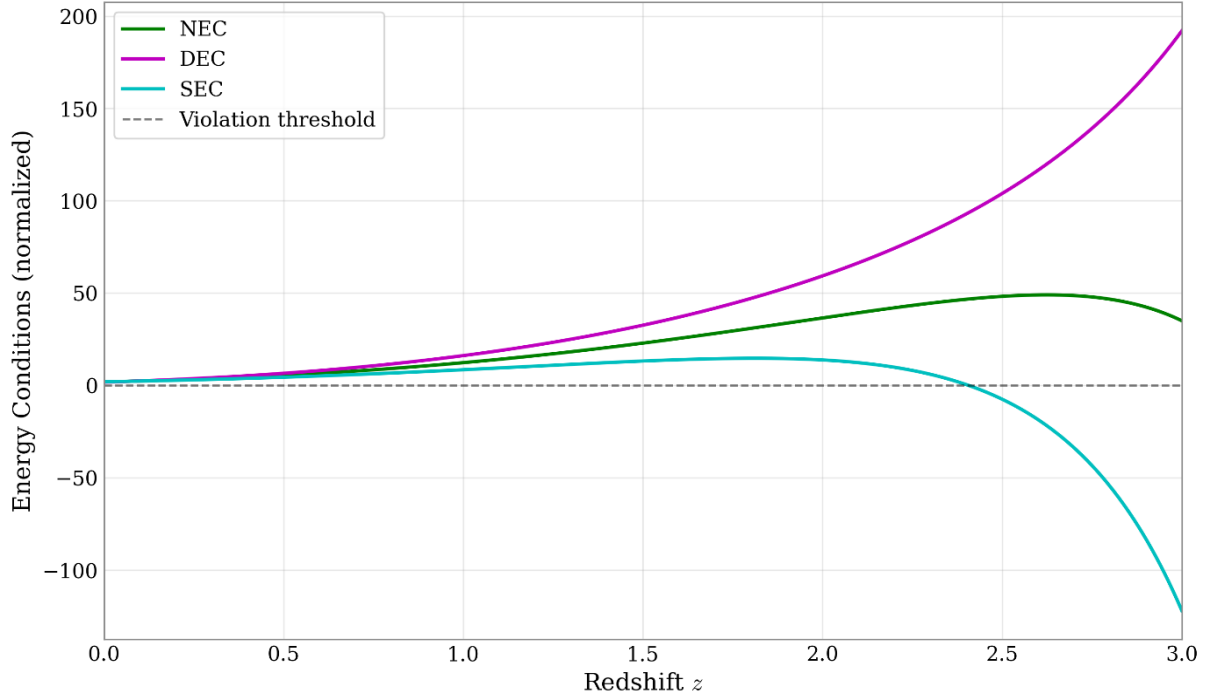

**Figure 10.** Evolution of the normalized energy conditions (NEC, DEC, and SEC) as functions of redshift $z$ for the square model. The dashed line indicates the violation threshold.

## 5.3 The Power-Law Model

Finally, to allow for greater departures from linearity and to explore a broader range of cosmological behaviors, we generalize the torsion contribution by introducing a power-law dependence while keeping the boundary term linear. This leads to the Power-law model:

$$f(T,B) = T + \varepsilon(-T)^n + \mu B, \qquad (45)$$

where $n$ and $\mu$ both dimensionless parameters, and $\varepsilon = \tilde{\varepsilon}/M^{2(n-1)}$, (with $M$ being a mass scale). This functional form is motivated by analytic and power-law reconstruction approaches in recent $f(T,B)$ studies [16,47], which show that non-linear torsion contributions can capture a wide range of cosmological behaviors. Although similar forms have been studied, the present ansatz — a power-law in $T$ combined with a linear boundary term — enables the exploration of cosmological dynamics not previously addressed, in both early- and late-time evolution.

This model includes several notable limits. Sitting $\mu = 0$, one obtains a pure power-law $f(T)$ theory, which for $n = 1$ reduces to a simple rescaling of the TEGR term with $G_{eff} = G/(1-\varepsilon)$. Conversely, setting $\varepsilon = 0$ and $\mu = -1$ yields $f(T,B) = -R$, thereby reproducing GR. One can obtain $\varepsilon$ by plugging (Eq. 45) in (Eq. 4) at the present time.

$$\varepsilon = \frac{(\Omega_{m0} + \Omega_{r0}) + 5\mu + 3}{(2n+1)(6H_0^2)^{n-1}}, \qquad (46)$$

Substituting (Eq. 45) in (Eq. 7), one obtains $\rho_{DE}$ as:

$$\rho_{DE} = \rho_r \left[4 + 2\mu - (\Omega_{m0} + \Omega_{r0} + 5\mu + 3)\left(\frac{H}{H_0}\right)^{2(n-1)}\right], \quad (47)$$

Plugging $\rho_{DE}$ (Eq. 47), and (Eq. 10) into (Eq. 15), we have:

$$\frac{\delta\mathcal{T}_f}{\mathcal{T}_f} = \sqrt{\frac{\pi^2 g_*}{90}\left[4 + 2\mu - (\Omega_{m0} + \Omega_{r0} + 5\mu + 3)\left(\frac{\mathcal{T}_f}{\mathcal{T}_0}\right)^{4n-4}\right]} \frac{1}{10qM_P\mathcal{T}_f^3}, \quad (48)$$

For $\mu = -2$, the model, as shown in Fig. 11, intersects the BBN exclusion bound at $n \approx 0.92$. Thus, consistency with BBN observations requires $n \leq 0.92$.

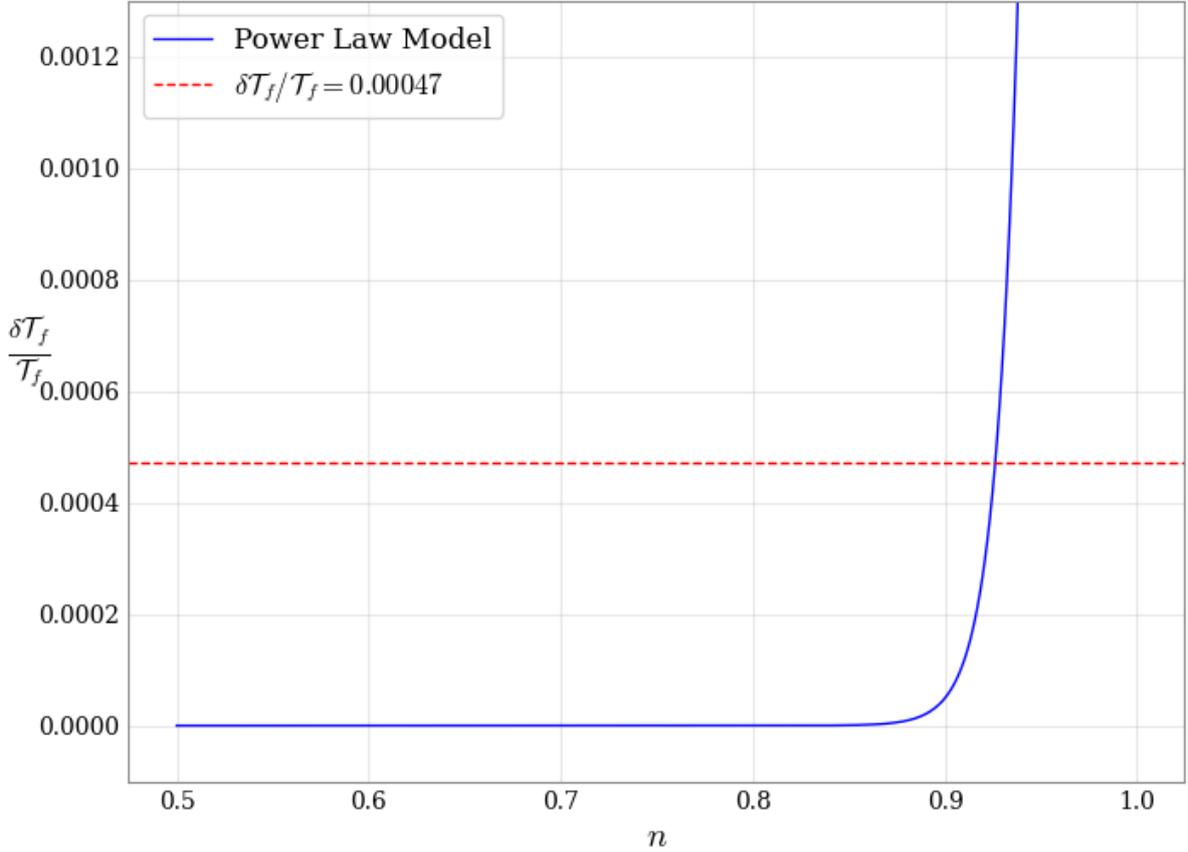

**Figure 11.** $\delta\mathcal{T}_f/\mathcal{T}_f$ vs. $n$ for $\mu = -2$; the BBN bound intersects the model prediction at $n \approx 0.92$, enforcing $n \leq 0.92$ to be observationally acceptable.

By substituting the functional form of the power law model along with (Eq. 46) into (Eq.4), we derive

$$\frac{dE(z)}{dz} = \frac{E(z)}{2(1+z)\mu}\left[(\Omega_{m0}(1+z)^3 + \Omega_{r0}(1+z)^4)E^{-2}(z) + 6\mu + 3 \\ - \left((\Omega_{m0} + \Omega_{r0}) + 5\mu + 3\right)E^{2n-2}(z)\right], \quad (49)$$

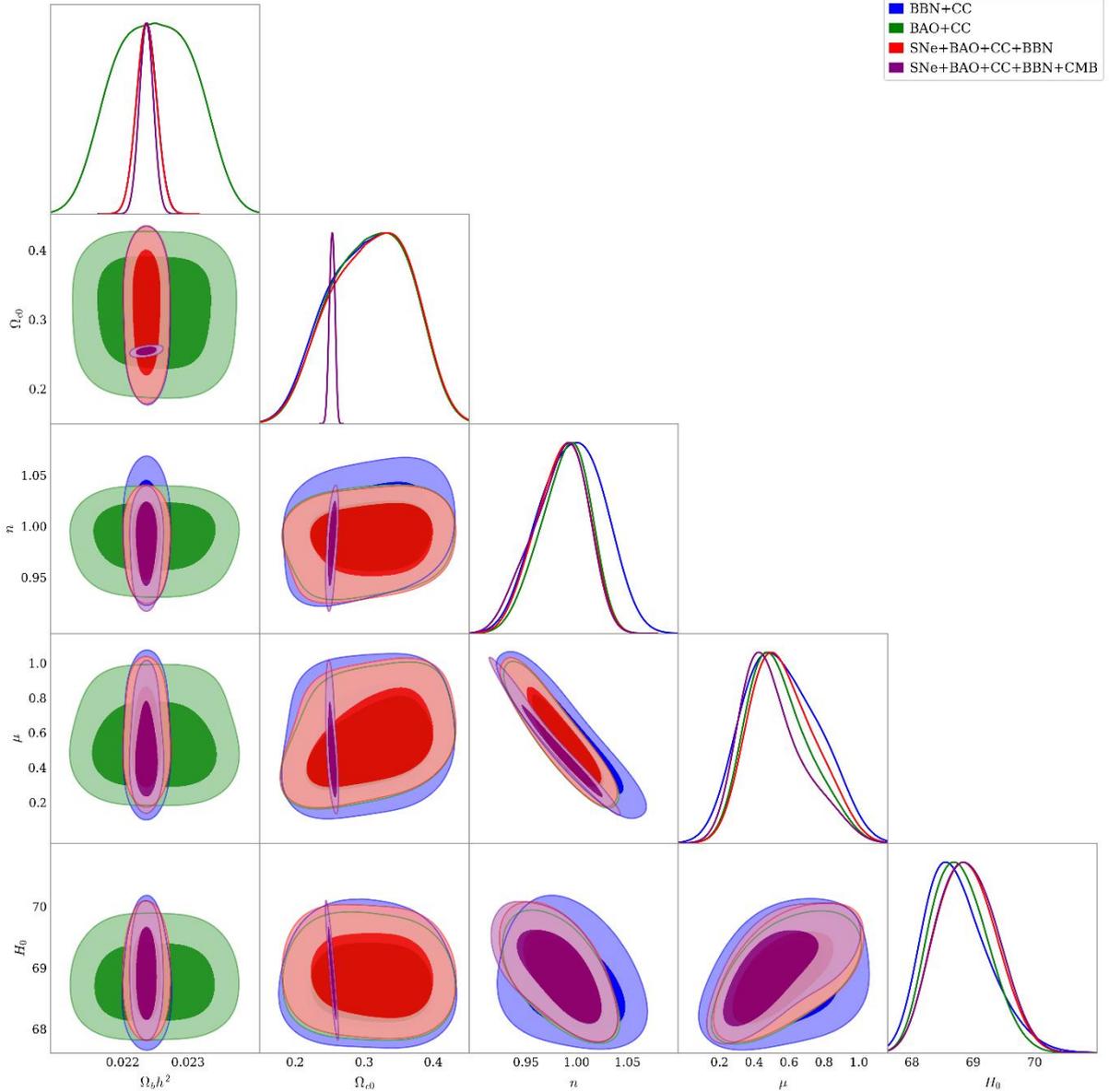

**Figure 12.** Corner plot illustrating the joint posterior distributions of the power-law model parameters for different observational datasets, showing 68% and 95% confidence regions.

Table 3 lists the cosmological constraints for the power law model. Fig. 12 shows the corner plot for the joint posterior distributions of the power-law parameters for different observational datasets, whereas individual independent contributions from CC (SNe Ia) are shown in Fig. 13(14). We used (Eq. 49) in producing these figures.

We see that $\Omega_{m0}$ ranges from $0.30 - 0.36$, whereas the modified gravity parameter $n$ has a larger uncertainty in individual datasets (typically (n > 0.95)) while it narrows to $n \approx 0.988$ for the combined fit. The parameter $\mu$ decreases from $\sim 0.61$ (SNe Ia) to $\sim 0.46$ when all datasets are combined, indicating convergence under common constraints.

**Table 3.** Best-fitting parameters for the power-law model from different cosmological data sets, with 68% confidence intervals and $\chi_\nu^2$ values.

| Dataset | $\Omega_{m0}$ | $\mu$ | $n$ | $H_0$ | $\chi^2_\nu$ |
|---|---|---|---|---|---|
| SNe | $0.3468^{+0.0701}_{-0.0695}$ | $0.6120^{+0.2737}_{-0.3062}$ | $0.9527^{+0.2896}_{-0.2589}$ | $71.03^{+1.35}_{-1.72}$ | 0.2 |
| CC | $0.3599^{+0.0631}_{-0.0730}$ | $0.5553^{+0.2799}_{-0.2101}$ | $0.9932^{+0.0334}_{-0.0347}$ | $68.68^{+0.61}_{-0.46}$ | 0.55 |
| BBN+CC | $0.3613^{+0.0601}_{-0.0724}$ | $0.5325^{+0.2784}_{-0.1972}$ | $0.9974^{+0.0316}_{-0.0356}$ | $68.65^{+0.61}_{-0.45}$ | 0.54 |
| BAO+CC | $0.3616^{+0.0600}_{-0.0685}$ | $0.5050^{+0.2300}_{-0.1409}$ | $0.9922^{+0.0207}_{-0.0293}$ | $68.75^{+0.50}_{-0.45}$ | 0.76 |
| SNe+BAO+CC+BBN | $0.3627^{+0.0597}_{-0.0707}$ | $0.5377^{+0.2383}_{-0.1561}$ | $0.9882^{+0.0226}_{-0.0291}$ | $68.88^{+0.51}_{-0.50}$ | 0.22 |
| SNe+BAO+CC+BBN+CMB | $0.3013^{+0.0031}_{-0.0030}$ | $0.4555^{+0.2435}_{-0.1217}$ | $0.9881^{+0.0215}_{-0.0344}$ | $68.89^{+0.53}_{-0.49}$ | 0.22 |

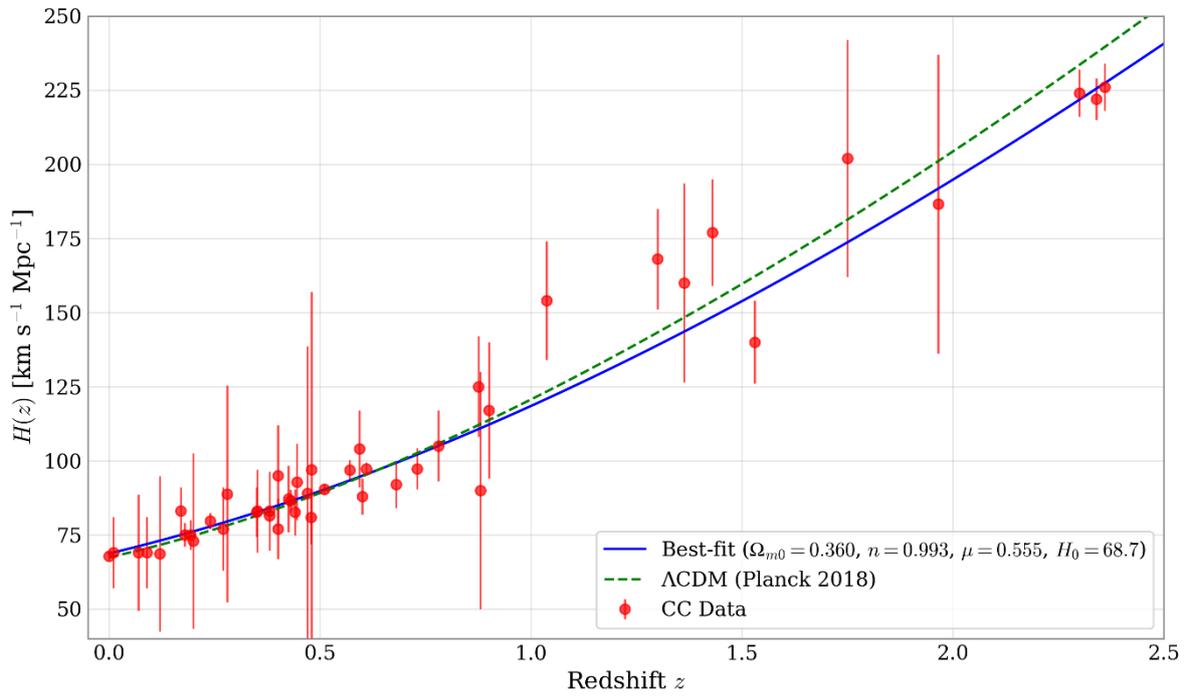

**Figure 13.** Reconstructed evolution $H(z)$ for the power-law model using CC data. The solid line represents the best-fit model, while the dashed line corresponds to the ΛCDM prediction from Planck 2018.

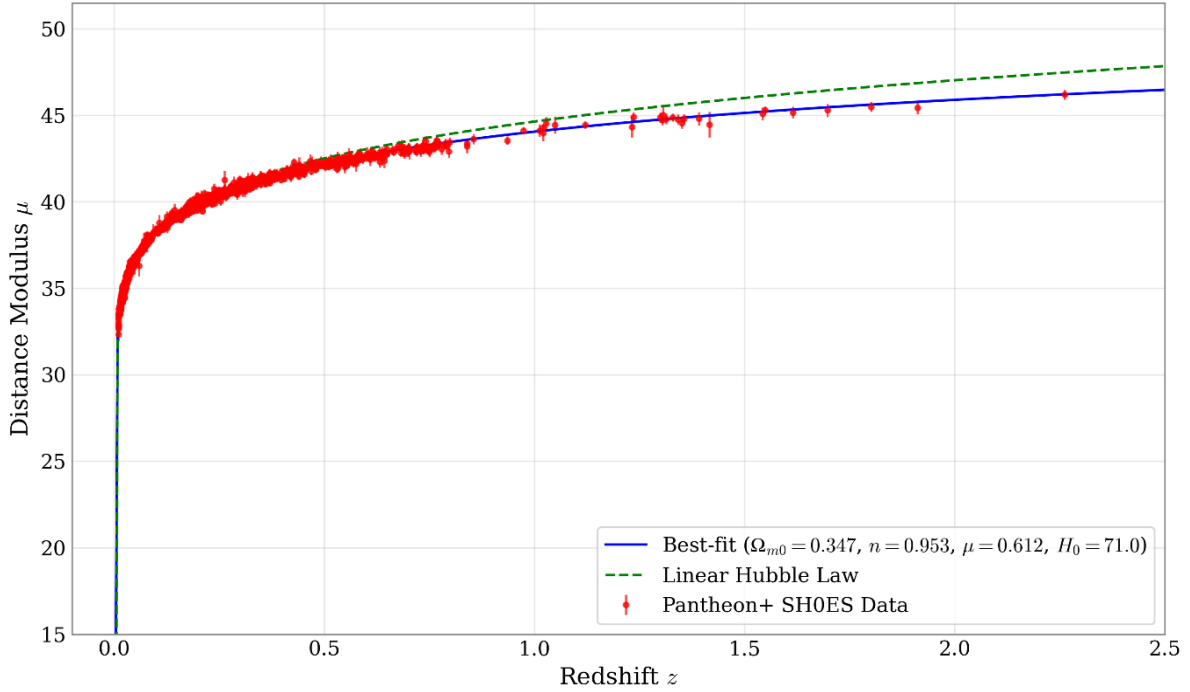

**Figure 14.** Distance modulus–redshift relation for the power-law model fitted to Pantheon+SH0ES Type Ia supernova data. The solid curve denotes the best-fit model, while the dashed line shows the linear Hubble law.

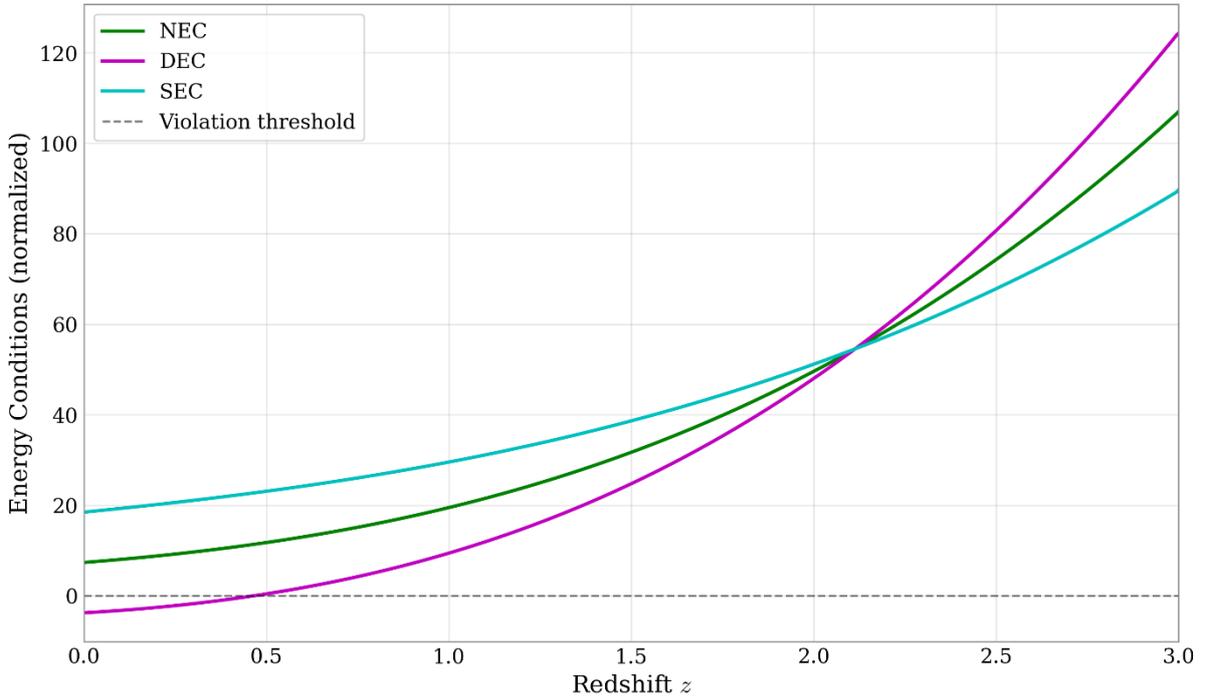

**Figure 15.** Evolution of the normalized energy conditions (NEC, DEC, and SEC) as functions of redshift $z$ for the power-law model. The dashed line marks the threshold of violation.

The helium limit requires $n \leq 0.92$, for fixed $\mu = -2$, which is inconsistent with the late data. Fixing the parameter $\mu = 0.35$, a value taken in line with the joint fittings in Table 3, will yield in the joint fit the value $n = 1.009$, restoring agreement between the early and late data. The Hubble parameter inferred from SNe is $\sim 71$ (km s$^{-1}$Mpc$^{-1}$), whereas the other probes favor a value closer to $\sim 68$ (km s$^{-1}$Mpc$^{-1}$). For the combined case (SNe Ia + BAO + CC + BBN + CMB), the energy-state results (Fig. 15) show that NEC and SEC are satisfied, while DEC is violated—indicating a non-standard but physically realistic energy-momentum behavior associated with late-time cosmic acceleration.

For values constrained by BBN: $\mu = 0.35$ and $n = 0.98$, and with $(M = H_0)$, the dimensionless parameter $\tilde{\varepsilon}$ can be computed from the analytic expression:

$$\tilde{\varepsilon} = \frac{\Omega_{F0} - 5\mu - 4}{(2n-1)6^{1-n}}, \qquad (50)$$

giving, when substituting the parameter values into (Eq. 50), $\tilde{\varepsilon} \approx -5.061$.

# 6 Conclusion

In this work, we explore three typical gravitational models for $f(T, B)$ gravity —linear, square, and power-law—over a wide range by combining BBN limits in the early Universe with late-time cosmological observations, including SNe Ia, BAO, CC, and CMB observations. In addition to comparing observations with these models predictions, we also examine the physical feasibility of the models by examining key energy conditions (NEC, DEC, and SEC). By considering early and late epochs, this approach allows us to analyze how consistent each model is with the overall history of cosmic evolution.

Our results show that the linear model fits the late-time cosmological data very well, with the estimated parameters being very close to those of ΛCDM: $\Omega_{m0} \approx 0.303 \pm 0.003$ and $H_0 \approx 68.7$ km s$^{-1}$Mpc$^{-1}$. However, the model does not meet the BBN requirements because the parameter ($\gamma$) falls outside the acceptable region for helium abundance ($0.229 \leq \gamma \leq 0.237$). Although the linear model satisfies important energy conditions, it cannot simulate the physical environment of the early universe. In contrast, the square model has remarkable homogeneity in both early and late observations. Its parameter $\lambda = 0.5312 \pm 0.0015$ falls comfortably within the BBN estimate $\lambda = 0.5369$, and $\Omega_{m0}$ and $H_0$ fall within the cosmological ranges. Energy state analysis finds that NEC and DEC are satisfied, and a very small violation of SEC—comfortably lying within the observed acceleration in the late universe. The power-law model, with a coupling coefficient of $\mu = 0.35$, yields $n \approx 1.009$, which is compatible with both the BBN and late-time observation requirements. The former, however, has a tension: late observations favor $n > 0.95$, while BBN requires $n \leq 0.92$. The model satisfies the NEC and SEC criteria but violates the WEC criteria, namely, non-standard effective energy and momentum behavior, which could be related to late-time cosmic acceleration.

While all three $f(T, B)$ models considered here closely reproduce the background expansion history and late-time cosmological data, certain features could distinguish them from ΛCDM. The Power-law model exhibits WEC violations, potentially affecting late-time cosmic acceleration, while the Square model shows minor SEC deviations that may subtly alter the deceleration-to-acceleration transition. Moreover, all models predict modified growth of dark matter perturbations and matter power spectra relative to ΛCDM, which could be probed via redshift-space distortions (RSD), lensing, and clustering observations. These deviations, particularly at the level of structure formation and early-universe constraints (BBN), represent possible observational signatures of $f(T,B)$ gravity.

Overall, the three models provide a very accurate simulation of cosmological observations, as reflected in their low chi-square values, while generally displaying physically plausible behavior. Among these models, the square and power-law models appear particularly promising, providing self-consistent descriptions of the evolution of the early and late universe. We focused on models with non-linearity primarily in $T$ because they are well-motivated as direct extensions of $f(T)$ gravity, which has been extensively studied in cosmology. Subsequent investigations could extend this analysis to more general functions $f(T,B)$.

Actually, from an effective field theory (EFT) perspective, $T$ and $B$ can be viewed as low-energy operators that encode the torsional and boundary contributions to the gravitational action. In the teleparallel framework, $T$ is the leading-order torsion scalar, analogous to the Ricci scalar in curvature-based EFT expansions, while $B$ acts as a boundary term that connects $T$ to the Ricci scalar via $R = -T + B$. In constructing our models, we prioritized non-linearity in $T$ because it represents the dominant torsional operator at late times and can be naturally motivated as a power-series expansion, where higher powers of $T$ capture leading-order corrections beyond TEGR. In this perspective, our chosen forms—linear in $T$ with constant affine terms and power-law in $T$—can be seen as the first non-trivial terms in a Taylor expansion of a more general $f(T,B)$ function, where $B$-dependence is subdominant or absorbed into redefined coefficients at the energy scales relevant to cosmology.

A natural extension is to include nonlinear $B$-terms—e.g., $f(T,B) = T + \varepsilon T^n + \gamma B^m$—guided by an EFT expansion. Exploring such generalized forms, alongside a careful stability assessment to check for potential ghost-like degrees of freedom, represents a compelling direction for subsequent studies.

Moreover, additional observational probes—such as gravitational baryon composition or higher-resolution CMB anisotropy measurements—could further constrain torsion-related couplings and test the consistency of these alternative gravity theories. In particular, although we focused on background evolution to establish consistency with cosmological observations, but perturbation analyses are necessary to study structure formation, since modified effective gravity in $f(T,B)$ can alter the growth of dark-matter perturbations and the matter power spectrum. Confronting these predictions with RSD, lensing, and clustering data constitutes a key avenue for future research.

# Acknowledgments

This research received no specific grant from any funding agencies in the public, commercial, or not-for-profit sectors. The author acknowledges the use of publicly accessible cosmological data from the Planck 2018 Legacy Archive and preprints hosted on arXiv, which were indispensable to this work. Gratitude is extended to the developers of the open-source Python libraries NumPy, SciPy, and Matplotlib, whose tools were critical to the analysis, numerical computations, and visualization of results presented in this study. N.C. acknowledges support from the PIFI-CAS fellowship and from Humboldt Foundation.

**ORCID iD**

**Yahia Al-Omar**    https://orcid.org/0009-0009-0045-4129

**Majida Nahili**    https://orcid.org/0000-0003-0103-2146

**Nidal Chamoun**    https://orcid.org/0000-0002-6140-643X